\documentclass[%
 reprint,
 prd,
nofootinbib,
 amsmath,amssymb,
 aps,
 lengthcheck,
floatfix,
]{revtex4}

\usepackage{physics}
\usepackage{graphicx}
\usepackage{dcolumn}
\usepackage{bm}
\usepackage{hyperref}
\usepackage[utf8]{inputenc}
\usepackage{multirow}
\usepackage{soul}
\usepackage{float}
\usepackage{graphicx}
\usepackage{times}
\usepackage[normalem]{ulem}
\usepackage{color}
\usepackage{cellspace}
\usepackage{lipsum, babel}
\usepackage{lipsum}
\usepackage{xcolor}
\usepackage{orcidlink}
\definecolor{customviolet}{RGB}{148,0,211}
\newcommand{\be}{\begin{equation}}
\newcommand{\ee}{\end{equation}}
\newcommand{\bea}{\begin{eqnarray}}
\newcommand{\eea}{\end{eqnarray}}

\newcommand{\comment}[1]{}
\renewcommand\sout{\bgroup \color{red} \ULdepth=-.5ex \ULset}
\def\simge{\mathrel{\rlap{\raise 0.511ex
     \hbox{$>$}}{\lower 0.511ex \hbox{$\sim$}}}}
\def\simle{\mathrel{\rlap{\raise 0.511ex
      \hbox{$<$}}{\lower 0.511ex \hbox{$\sim$}}}}

\usepackage{multirow}

\DeclareUnicodeCharacter{2212}{-}

\begin{document}


\title{Realizing the potential of deep neural network for analyzing neutron star observables and dense matter equation of state} 

\author{Ameya Thete \orcidlink{0000-0002-8089-5945}}
\email{f20180885@goa.bits-pilani.ac.in}
\affiliation{Department of Physics, BITS-Pilani, K. K. Birla Goa Campus, Goa 403726, India}

\author{Kinjal Banerjee \orcidlink{0000-0002-8864-4120}}
\email{kinjalb@gmail.co}
\affiliation{Department of Physics, BITS-Pilani, K. K. Birla Goa Campus, Goa 403726, India}

\author{Tuhin Malik \orcidlink{0000-0003-2633-5821}}
\email{tuhin.malik@uc.pt}
\affiliation{CFisUC, Department of Physics, University of Coimbra, P-3004 - 516  Coimbra, Portugal}
\date{\today}

\begin{abstract} 
The difficulty in describing the equation of state (EoS) for nuclear matter at densities above the saturation density ($\rho_0$) has led to the emergence of a multitude of models based on different assumptions and techniques. These EoSs, when used to describe a neutron star (NS), lead to differing values of observables. An outstanding goal in astrophysics is to constrain the dense matter EoS by exploiting astrophysical and gravitational wave measurements. Nuclear matter parameters appear as Taylor coefficients in the expansion of the EoS around the saturation density of symmetric and asymmetric nuclear matter and provide a physically-motivated representation of the EoS. In this paper, we introduce a deep learning-based methodology to predict key neutron stars observables such as the NS mass, NS radius, and tidal deformability from a set of nuclear matter parameters. Using generated mock data, we confirm that the neural network model is able to accurately capture the underlying physics of finite nuclei and replicate inter-correlations between the symmetry energy slope, its curvature, and the tidal deformability arising from a set of physical constraints. We also test our network with mock data generated by a different class of physics model, which was not part of the training, to explore the limitations of model dependency in the results. We also study the validity of our trained model using Bayesian inference and show that the performance of our model is on par with physics-based models with the added benefit of much lower computational cost.
\end{abstract}


\maketitle

\section{Introduction}
Neutron Stars (NSs) are some of the most fascinating astrophysical objects in multi-messenger astronomy. They contain matter at extreme conditions far beyond the ones accessible in a terrestrial laboratory. The core of a NS is believed to contain matter at a few times nuclear saturation density, $\rho_0$  \cite{Glendenning1996,Haensel2007,Most:2018hfd,Ruester:2005fm}\footnote{$\rho_0$ = 0.16 fm$^{-3}$}. While the structure of a NS can be determined using the Tolman-Oppenheimer-Volkoff (TOV) equations \cite{TOV1,TOV2}, this requires the knowledge of the dense matter Equation of State (EoS). Understanding the internal structure of neutron stars in terms of fundamental interactions between its constituents is an open problem in nuclear physics. To understand the physics of dense matter, at low densities of $\rho \sim 1-2\rho_0$, we can use \textit{ab initio} approaches derived from chiral effective field theory ($\chi$EFT) \cite{Hebeler:2009iv, Tews:2012fj, Hagen:2013yba, Roggero:2014lga, Krastev:2021reh}. At large densities of $\rho \geq 40\rho_0$, perturbative quantum chromodynamics (QCD) calculations converge and provide reliable estimates \cite{Freedman:1976xs, Kurkela:2009gj, Gorda:2018gpy, Krastev:2021reh}. However, in the intermediate density region around $\rho \sim 2-10\rho_0$ which is relevant for most structural descriptions of neutron stars, reliable calculations from first principles are currently unavailable \cite{Krastev:2021reh, Fujimoto:2021zas}. Quantum field theory calculations based on lattice QCD are challenging at these densities due to the sign problem that arises in Monte Carlo simulations \cite{Aarts:2015tyj}. As a result, structural descriptions of neutron stars rely on relativistic and non-relativistic phenomenological models for the EoS. The nuclear matter parameters (NMPs), which form the basis of the construction of these equations of state for neutron star matter, are not directly accessible. While lower-order NMPs can be empirically extracted through finite nuclei nuclear physics experiments \cite{Lalazissis:1996rd, Todd-Rutel:2005yzo, Klupfel:2008af, Sulaksono:2009rn}, in order to constrain higher-order NMPs, we need to rely on astrophysical observations \cite{Zhang:2018vrx, Cai:2020hkk, Gil:2020wqs}.

Recent developments in multi-messenger astronomy have provided important information about high-density nuclear matter physics relevant for NSs. Constraining the EoS is a joint task between nuclear physics and astrophysics. Measured astrophysical quantities such as NS observables can uncover properties of dense nuclear matter. Narrowing the constraints on NS observables, therefore, has the ability to constrain the behavior of matter under extreme conditions. It is expected that precise and simultaneous measurements of NS properties like mass, radius, moment of inertia and tidal deformability may help constrain the EoS to a narrow range \cite{Annala:2021gom, Altiparmak:2022bke, Margueron:2017eqc, Margueron:2017lup}. High mass pulsars like PSR J1614--2230 ($M = 1.908 \pm~ 0.016 M_{\odot}$) \cite{Demorest:2010bx, Fonseca:2016tux, NANOGrav:2017wvv}, PSR J0348--0432 ($M = 2.01 \pm~ 0.04~ M_{\odot}$) \cite{Antoniadis:2013pzd},  PSR J0740+6620 ($M = 2.07 \pm~ 0.07~ M_{\odot}$  \cite{Fonseca:2021wxt}, and very recently PSR J1810+1714 ($M = 2.13 \pm~ 0.04~ M_{\odot}$) \cite{Romani:2021xmb} have already placed tight constraints on the EoS. GW signals emitted from a NS merger event depend on the behavior of the neutron star matter at high densities \cite{Faber:2009zz, Duez:2009yz}. Therefore, the values of tidal deformability obtained from GW events such as GW170817 associated with a binary NS merger, as well as the simultaneous measurement of NS masses and radii from high-precision X-ray space missions, such as NICER (Neutron star Interior Composition ExploreR), may help further constrain the EoS. Some current observational evidences are the simultaneous measurements of NS mass $1.34_{-0.16}^{+0.15}$ $M_\odot$ and radius $12.71_{-1.19}^{+1.14}$ km for the pulsar PSR J0030+0451 by NICER \cite{Riley:2019yda}.  Other independent analyses show that the radius is $13.02_{-1.06}^{+1.24}$ km and the mass $1.44_{-0.14}^{+0.15}$ $M_\odot$ \cite{Miller:2019cac}. However the recent measurement of the equatorial circumferential radius of the  pulsar PSR J0740+6620 with  mass $M=2.072_{-0.066}^{+0.067}$ $M_\odot$ and   $R=12.39^{+1.30}_{-0.98}$ km (68 $\%$ CI) \cite{Riley:2021pdl} cannot further constrain the EoS which already predicts a NS with maximum mass more than 2$M_\odot$ \cite{Malik:2022zol}.

The EoS -- to a good approximation -- can be expressed in terms of NMPs at saturation density. The NMPs usually considered for constructing the EoS are the incompressibility coefficient, the skewness parameter of the symmetric nuclear matter, the symmetry energy coefficient, its slope, and the curvature parameters characterizing the density dependence of the symmetry energy. Recently, there has been a comprehensive analysis of correlations of tidal deformability and other NS properties with NMPs \cite{Fattoyev:2017jql, Zhang:2018vrx, Carson:2018xri}; however, these correlations are found to be model-dependent \cite{Carson:2018xri,Ghosh:2022lam}. Ref. \cite{Malik:2020vwo} shows that the correlations are sensitive to finite nuclei properties, which are accessible to laboratories. Therefore, finite nuclei properties are important quantities to consider in a model while determining the correlations between NS properties and NMPs. Of late, the EoSs obtained by several meta-models have gained popularity owing to their cost-effectiveness in big simulations \cite{Annala:2021gom, Annala:2019puf, Kurkela:2009gj}. These models are constrained by {\it ab initio} theoretical calculations of nucleon-nucleon chiral potentials for low-density neutrons and nuclear matter \cite{Hebeler:2013nza,Drischler:2015eba},  and  perturbative QCD for asymptotically high-density regimes \cite{Kurkela:2009gj}. In the intermediate density region, the EoS is evolved in a thermodynamically consistent manner with either piece-wise polytropic segments \cite{Read:2008iy, Ozel:2015fia, Raithel:2017ity}, a speed-of-sound interpolation, or a spectral interpolation \cite{Lindblom:2010bb, Lindblom:2012zi}. These meta-models are limited to incorporating finite nuclei properties and differ on results in establishing a bridge between NS properties and NMPs. Non-parametric models of the NS EoS have also been proposed based on Gaussian processes (GPs) \cite{Essick:2019ldf, Landry:2020vaw} which use Bayesian methods to infer the EoS from multi-messenger data. These models are usually computationally expensive to implement or are highly sensitive to the choice of the training data sets and therefore, might be limited by the current knowledge of the EoS \cite{Han:2021kjx}. Consequently, there is a need to search for alternative approaches to construct a model-independent EoS. 

In recent years, deep learning (DL) has been extensively applied to a wide range of technological and scientific tasks. DL algorithms, which are a class of machine learning (ML) algorithms, are highly scalable and distributed computational techniques with the ability to learn intricate relationships from raw data using units called neurons arranged in a stacked fashion. The advent of high-performance computing and the development of parallel devices like graphics processing units have rendered DL as the primary choice of algorithms for tasks such as computer vision \cite{He2016}, or natural language processing \cite{Young2018}.  DL models have been used as alternatives to conventional statistical framework and have been successfully applied to many problems in physics. Most applications of ML and DL in physics have been in analyzing data obtained from data-intensive experiments like LIGO for the detection and denoising of GW signals \cite{George:2016hay, George:2017pmj, Carrillo:2016kvt}, and the Large Hadron Collider for particle track reconstruction or anomaly detection \cite{Larkoski:2017jix, Guest:2018yhq, Bourilkov:2019yoi}. Significant progress has also been made in using these algorithms in the context of nuclear physics \cite{bedaque2020} and neutron star physics \cite{Fujimoto:2019hxv, Han:2021kjx, Morawski:2020izm}. While these applications are certainly promising, it remains to be seen up to what extent a DL model can supplement existing physical models. Recent research also aims to address whether a trained DL model can produce correct predictions from experimental data alone and whether such predictions are comparable to mesoscopic phenomenological physics models \cite{Anil:2020lch}. Such ML and DL-based models do not have the feature richness possessed by physics-based models but they offer other benefits like cost-effectiveness while dealing with a large amount of experimental or observational data. Recent works \cite{Fujimoto:2017cdo, Fujimoto:2019hxv, Fujimoto:2021zas} have studied the applications of machine learning methods to the neutron star EoS. They employ a feedforward neural network (FFNN) to map neutron star data to EoS parameters. Instead of considering the FFNN as merely an interpolation tool between EoS NMPs and neutron star observables, we adopt an approach similar to \cite{Han:2021kjx} by treating the FFNN as a representation of the EoS itself. 

{In this paper, our focus is on implementing an artificial neural network to emulate a physics model and assess its accuracy through validation and inference tasks. Our primary goal is to thoroughly examine the neural network's ability to mimic the physics model, and subsequently utilize the trained network for computationally expensive Bayesian inference tasks.  To the best of our knowledge, the use of neural networks in Bayesian inference framework has not been employed in nuclear astrophysics literature. We show that we can indeed obtain results comparable to physics models at a fraction of the computational cost. This points out a novel utilization of machine learning and neural network models in research. Let us however clarify that we are not claiming that neural networks can replace physics based models. Our claim is the such models can help us capture some of the features of the EoS at much lower computational costs.}

The paper is structured as follows: in Section \ref{sec:framework}, we discuss the parameterization of the EoS which leads to the emergence of NMPs, followed by a brief review of the theory of artificial neural networks and the DL approach used to map NMPs to NS observables. This is followed by a description of the Bayesian statistical framework. {In \ref{sec:results} we test how well does the best fit neural network model capture correlations between EoS parameters. We also perform a Bayesian analysis to compare our models with a physics based model and show that the computational cost is much less for our model. We finally end with our conclusions in our Section \ref{sec:conclusion}.}

\section{Framework}
\label{sec:framework}
In this section, we outline the different facets of the adopted framework for our analysis. In Sec. \ref{sec:nmps}, we briefly describe the neutron star EoS and the nuclear matter parameters involved in its construction. Then, in Sec. \ref{sec:data}, we describe the generation of the data set used to train the neural network model. Sec. \ref{sec:anns} describes at length the DL approach used to construct a deep neural network model which accepts nuclear matter saturation parameters as inputs and produces neutron star properties obtained from a set realistic nuclear physics EoSs as targets. Finally, in Sec. \ref{sec:bayes} we present a Bayesian inference framework that is applied on the trained neural network model to verify the  performance of our model in comparison with Skryme models.

\subsection{Nuclear matter parameters}
\label{sec:nmps}
The structure of neutron stars is obtained by solving the Tolman-Oppenheimer-Volkoff (TOV) equation with a given EoS for the nuclear matter. The EoS is expressed as the variation of pressure $e$ with density $\rho$, over a wide range of densities. To a good approximation, any EoS calculated from phenomenological nuclear models can be decomposed into two parts, (i) the EoS for symmetric nuclear matter $e(\rho, 0)$; and (ii) a term involving the symmetry energy coefficient $S(\rho)$ and the asymmetry $\delta$,

\begin{equation}
 e(\rho,\delta) \simeq e(\rho,0) + S(\rho)\delta^2,
 \label{eq:eden}
\end{equation} 
where $\rho  = \rho_n + \rho_p$ is the baryon density, $\rho_n$ and $\rho_p$ are the neutron and proton densities respectively, and the asymmetry $\delta = (\rho_n - \rho_p)/\rho$. We can then characterize the density dependence of the energy density of symmetric matter around the saturation density $\rho_0$ in terms of a few bulk parameters by constructing a Taylor expansion around $\rho_0$. That is,
    \begin{equation}
        e(\rho, 0) = e_0 + \frac{K_0}{2}\left(\frac{\rho - \rho_0}{3\rho_0}\right)^2 + \frac{Q_0}{6}\left(\frac{\rho - \rho_0}{3\rho_0}\right)^3 + \mathcal{O}(4)
    \end{equation}
The coefficients $e_0, K_0, Q_0$ denote the energy per particle, the incompressibility coefficient, and the third derivative of symmetric matter at saturation density, respectively.
Similarly, the behavior of the symmetry energy around saturation can also be characterized in terms of a few bulk parameters, 
    \begin{equation}
        S(\rho) = J_{\rm sym,0} + L_{\rm sym,0}\left(\frac{\rho - \rho_0}{3\rho_0}\right) + \frac{K_{\rm sym, 0}}{2}\left(\frac{\rho - \rho_0}{3\rho_0}\right)^2 +  \mathcal{O}(3)
    \end{equation}
where $J_{\rm sym,0} = S(\rho_0)$ is the symmetry energy at saturation density. The incompressibility $K_0$, the skewness coefficient $Q_0$, the symmetry energy slope $L_{\rm sym,0}$, and its curvature $K_{\rm sym, 0}$ evaluated at saturation density, are defined in \cite{Vidana:2009is}. These quantities are the key nuclear matter parameters (NMPs) that describe any equation of state (EoS). Hence, an EoS can be represented by a point in the seven-dimensional parameter space of NMPs $\{e_0$, $\rho_0$, $K_0$, $Q_0$, $J_{\rm sym,0}$, $L_{\rm sym,0}$, and $K_{\rm sym, 0}\}$ \cite{Malik:2020vwo}. Symbolically, the $j^{\text{th}}$ EoS in this space is written as
\begin{align}
        \text{EoS}_{j} &= \{e_0, \rho_0, K_0, Q_0, J_{\rm sym,0}, L_{\rm sym,0}, \text{and } K_{\rm sym, 0}\}_{j} \nonumber\\
                       &\approx \mathcal{N}(\boldsymbol\mu, \boldsymbol\Sigma)
\end{align}
where $ \mathcal{N}(\boldsymbol\mu, \boldsymbol\Sigma)$ is a multivariate Gaussian distribution with $\boldsymbol\mu$ being the mean value of the nuclear matter parameters $\mathbf{p}$ and a covariance matrix $\boldsymbol\Sigma$. The diagonal elements of $\boldsymbol\Sigma$ represent the variance or the squared error for the parameters $p_i$. The off-diagonal elements of $\boldsymbol\Sigma$ are the covariances between different parameters $p_i$ and $p_j$, and denote the correlation coefficient between them. Hence, given a mean $\boldsymbol\mu$ and covariance matrix $\boldsymbol\Sigma$, a large number of EoSs can be obtained. 

While the Taylor expansion of the symmetric and asymmetric energy is only truly accurate around the saturation density, we can treat these expansions as a parameterization of the EoS --- similar to other adopted parameterizations --- with the condition that this representation asymptotically approaches the Taylor expansion in the limit $\rho \rightarrow \rho_0$ \cite{Zhang:2018vrx, Ferreira:2019bny}. This lets us ignore any issue arising with the convergence of the approximation. However, the higher-order NMPs obtained via this parameterization might markedly diverge from the actual nuclear matter expansion coefficients. They can be thought of as effective parameters that incorporate the effects of missing higher-order terms. Moreover, Refs. \cite{Margueron:2017eqc, Ferreira:2019bny} indicate that an EoS obtained from the Skyrme framework can be well-reproduced by considering Taylor coefficients until the third or fourth order. 

\subsection{Data set generation}
\label{sec:data}

\begin{figure}[ht]
    \centering
    \includegraphics[scale=0.4]{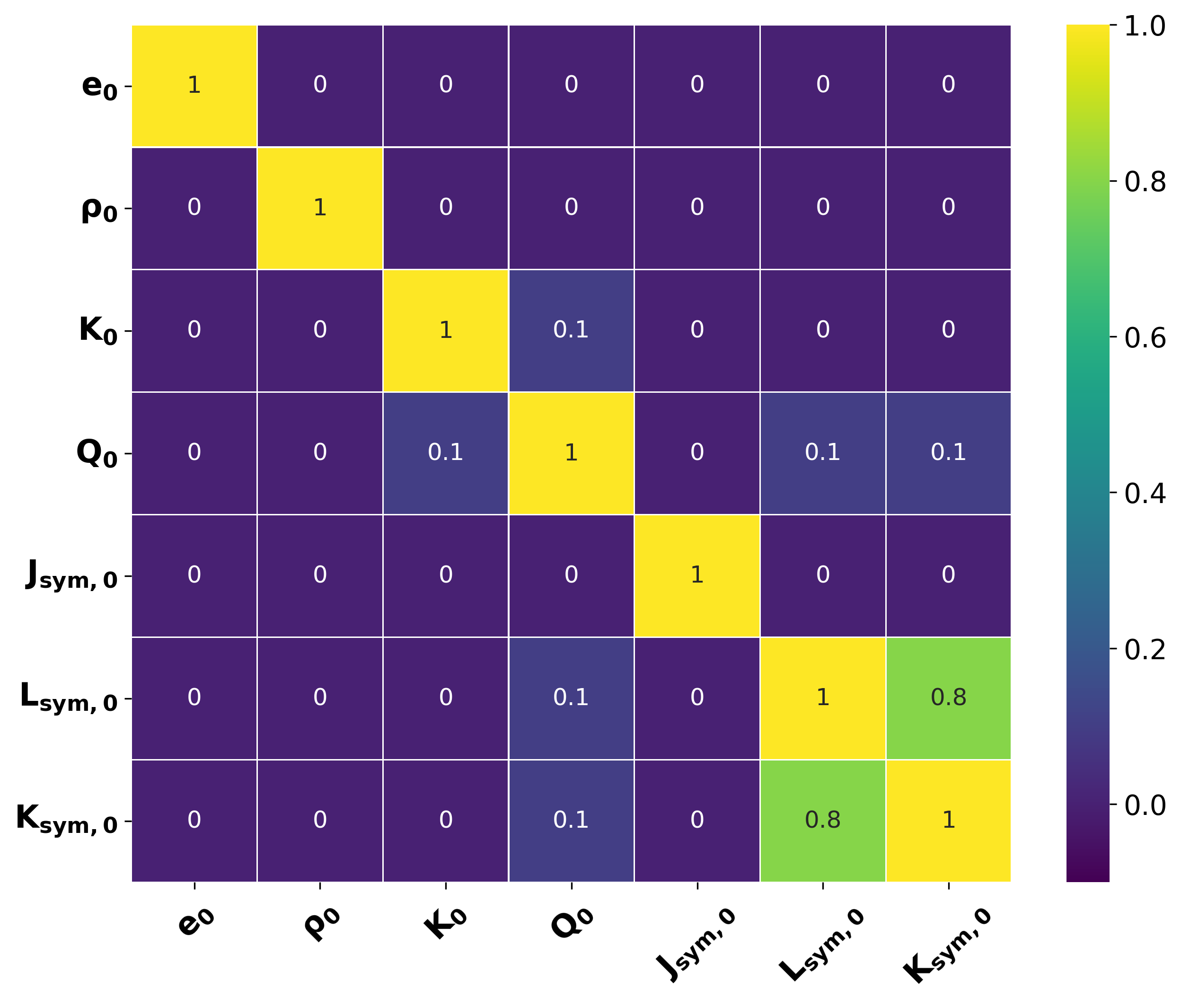}
    \caption{Correlations among the various NMPs for the sampled data set. Correlations among the off-diagonal pairs corresponding to $L_{\rm sym,0}-K_{\rm sym, 0}$
    are the only nontrivial values. The figure displays sample statistics that reveal insignificant correlations between other NMPs.}
    \label{fig:data-corr}
\end{figure}

In our analysis, the input data used to train an artificial neural network are the seven key nuclear matter parameters that govern the equation of state \{$e_0$, $\rho_0$, $K_0$, $Q_0$, $J_{\rm sym,0}$, $L_{\rm sym,0}$, $K_{\rm sym, 0}$\}. Six neutron star properties are the target variables: the maximum NS mass, $M_{\text{max}}$; the maximum NS radius, $R_{\text{max}}$; the radius for 1.4 $M_\odot$ NS, $R_{1.4}$; and the tidal deformability $\Lambda_M$ for NS having mass $M\in[1.0,1.4,1.8]M_{\odot}$. We generate our data set by sampling points from the multivariate Gaussian distribution $\mathcal{N}(\boldsymbol{\mu}, \boldsymbol{\Sigma})$, where $\boldsymbol\mu$ is the mean vector with components $\mu_i = \mathbb{E}[p_i]$ and $\boldsymbol\Sigma$ is the covariance matrix with entries $\Sigma_{ij} = \mathbb{E}[(p_i - \mu_i)(p_j - \mu_j)]$, for a NMP $p$  \footnote{$\mathbb{E}[X]$ is the expectation value of a random variable $X$ and is defined as $\mathbb{E}[X] = \int_{-\infty}^{\infty} xp(x)$, where $p(x)$ is the probability density function.}. This method closely follows the procedure for generating the \texttt{Case-II} data set in Ref. \cite{Malik:2020vwo}. We assume an a priori inter-correlation coefficient between $L_{\rm sym,0}$ and $K_{\rm sym,0}$ of 0.8, which is reasonable choice for nuclear physics models that satisfy finite nuclear properties. Models which satisfy finite nuclear properties exhibit different correlations between NMPs and NS properties as compared to meta-models or {nuclear physics model for infinite nuclear matter} that do not respect finite nuclear properties \cite{Carson:2018xri, Ghosh:2022lam}. Therefore, we consider \texttt{Case-II} data of Ref. \cite{Malik:2020vwo} to mimic the microphysics information of finite nuclear properties. Figure \ref{fig:data-corr} presents the $7 \times 7$ matrix for the correlation coefficients between the NMPs in the sampled data. The central values and uncertainties on each NMP in the constructed distribution are listed in Table \ref{tab:nmp-params}.

\begin{table}[h]
\centering
\begin{tabular}{ccccc}
\hline \hline 
NMP                & \multicolumn{2}{c}{MVGD}                                         & \multicolumn{2}{c}{Tranning Set}                                 \\ \hline
$\boldsymbol{p_i}$ & $\boldsymbol{\mu_{p_i}}$ & $\boldsymbol{\sqrt{\Sigma_{p_ip_i}}}$ & $\boldsymbol{\mu_{p_i}}$ & $\boldsymbol{\sqrt{\Sigma_{p_ip_i}}}$ \\ \hline
$e_0$              & $-16.0$                  & $0.25$                                & $-15.99$                 & $0.25$                                \\
$\rho_0$           & $0.16$                   & $0.005$                               & $0.158$                  & $0.005$                               \\
$K_0$              & $230$                    & $20$                                  & $242$                    & $15$                                  \\
$Q_0$              & $-300$                   & $100$                                 & $-307$                   & $70$                                  \\
$J_{\rm sym,0}$    & $32$                     & $3$                                   & $32.65$                  & $2.8$                                 \\
$L_{\rm sym,0}$    & $60$                     & $20$                                  & $73$                     & $12$                                  \\
$K_{\rm sym,0}$    & $-100$                   & $100$                                 & $-20$                    & $50$                                  \\ \hline
\end{tabular}
\caption{The mean value $\boldsymbol{\mu_{p_i}}$ and error $\boldsymbol{\sqrt{\Sigma_{p_i p_i}}}$ for the nuclear matter parameters $p_i$ employed for the 
        multivariate Gaussian distribution. All quantities are in units of MeV except for $\rho_0$ which is in units of fm$^{-3}$.}
        \label{tab:nmp-params}
\end{table}

We then construct the Skyrme EoS from a drawn NMP sample and check whether the EoS (a) predicts a NS maximum mass above 2 $M_\odot$; (b) predicts a tidal deformability for a NS with $M = 1.4 M_\odot$ below 800; and (c) satisfies the causality condition, i.e., the speed of sound $c_s = \sqrt{\dv{p}{e}} \leq c$ at the center of maximum mass NS, where $c$ is the speed of light in vacuum. Any samples that generate EoSs which do not satisfy these conditions are discarded. For each EoS, the TOV and deformability equations \cite{Hinderer:2007mb} are solved to obtain the six aforementioned NS observables. Naturally, realistic NS observations accrue experimental and instrumental errors, which result in corresponding uncertainties while reconstructing the NMPs. For simplicity, we choose to disregard NS observational errors and uncertainties while training the model. Following this procedure, we obtain 2106 filtered NMPs and corresponding NS observables which form the data set.  

\subsection{Artificial neural networks}
\label{sec:anns}

Artificial Neural Networks (ANN) are a class of machine learning algorithms that have found widespread use over the past decade. The popularity of ANNs arises from their ability to model complex nonlinear relationships within data and their potential to generalize to a wide class of functions. An ANN with sufficiently many layers or neurons is theoretically capable of representing any continuous function \cite{Hornik1989, Cybenko1989}. Therefore, as a general rule, a more complex network with a larger number of parameters is able to learn more abstract features from the data. 

\begin{figure}[pt]
    \centering
    \includegraphics{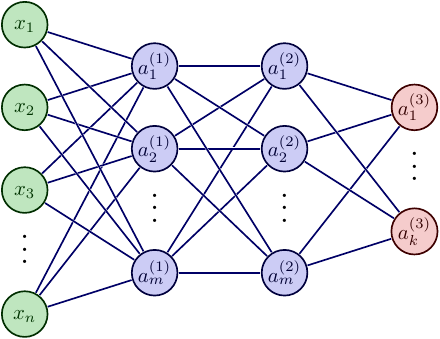}
    \caption{An example of a fully-connected feedforward neural network with two hidden layers. Each hidden layer consists of $m$ neurons, denoted by $a^{(l)}_{j}$, where $l$ denotes an index over the hidden layers and the index $j$ runs over the dimension of the hidden layer. This particular network accepts an $n$-dimensional input through the input layer (in green), passes the input through the hidden layers (in lavender), and produces a $k$-dimensional output at the output layer (in red). Solid lines between individual nodes 
    denote weighted connections that are learned during the training procedure. Activation functions are not shown in this figure.}
    \label{fig:ffnn-fig}
\end{figure}

Feedforward Neural Networks (FFNN) are the simplest type of neural network architecture. For an input $\vb{x}$, the objective of a feedforward neural network is to approximate some true mapping $f^*(\vb{x})$ by a mapping $f(\vb{x}; \boldsymbol\theta)$, parameterized by a set of weights $\boldsymbol\theta$. A typical feedforward neural network consists of a number of processing units called neurons, arranged into one or many layers composed in a sequential fashion. A neuron performs a linear operation by aggregating weighted inputs received from neurons in the previous layer. A FFNN generally consists of an input layer, followed by one or more hidden layers, and a final output layer consisting of one or more neurons. Computation in an FFNN flows in a linear fashion, starting at the input layer and moving successively through the hidden layers until it reaches the output layer. Figure \ref{fig:ffnn-fig} provides an illustration of a simple feedforward neural network with one input layer, two hidden layers and an output layer. 
    
The parameters $\boldsymbol\theta$ typically represent the weights assigned to a connection between neurons in adjacent layers. The training data provides noisy, approximate examples of $f^*(\vb{x})$ and each $\vb{x}$ is paired with a corresponding label $y$. The goal of a learning algorithm is to learn a particular value of $\boldsymbol\theta$ that results in the best function approximation. During the training procedure, the learning algorithm does not say what each layer does but instead decides how to use these layers to produce an optimal approximation of $f^*$. As the training data does not show the desired output for intermediate layers, they are called hidden layers. The number of hidden layers decides the depth of the network, and the dimensionality or the number of neurons in each hidden layer determines the width of the network. To introduce nonlinearity in the computation between two successive layers within the model, a nonlinear function, called the activation function, acts element-wise on the output of one hidden layer, and the output of the function is passed to the next layer in the computation. For most modern neural networks, the default recommendation is to use the rectified linear unit, or ReLU \cite{Jarrett2009, Nair2010} activation, defined as $g(z) = \max\{0, z\}$. Other commonly used choices for the activation function include the $\tanh$ function or the logistic function. For a textbook review of neural networks and training algorithms, see Ref. \cite{Goodfellow2016}.

The ability of neural networks to model highly nonlinear relationships between input and output variables makes them ideal for estimating neutron star properties from the equation of state parameters. This is important because the relationships between these two quantities are expected to be nontrivial and involve multiple intermediate steps composed of nonlinear operations. Moreover, a neural network offers two major advantages over a conventional approach with traditional physics models:
\begin{enumerate}
    \item  An ANN can efficiently map the sample NMPs to NS properties without calculating the EoS from nuclear physics models. Similar works \cite{Wei:2020xrl, Chua:2019wwt} demonstrate that ANNs offer up to a two-fold speedup over conventional astrophysical models
    \item ANNs can also accurately capture finite nuclear information, which can computationally expensive to verify with a traditional physics model in a Bayesian setting. 
\end{enumerate}
At this step, we wish to note that other machine learning models can also be used for an identical purpose, albeit with varying degrees of success. We performed a preliminary comparison of FFNNs with other ML models, specifically linear regression, support vector regression and eXtreme Gradient Boosting (XGBoost) \cite{xgboost}. We observed that FFNNs outperformed all the other models in the study, and therefore, we chose to use FFNNs for this work. 

In supervised learning instances, the data set is generally partitioned into non-overlapping training, testing, and validation sets. The training data set is used by the neural network to learn, the validation set is used to see whether the network is learning properly during the training procedure and to select the optimal values for the hyperparameters, and the testing data set is used to assess the performance of the trained model. It should be noted that the model does not see any instances from the testing data set until after the training is completed. Accordingly, we randomly partition the original generated data set following a 70\%-20\%-10\% split into a training set, a validation set, and a testing set consisting of 1515, 422, and 169 instances, respectively. Before training, the features in the split datasets are standardized by removing the mean and scaling them to unit variance. Standardization of data is a common requirement for many ML algorithms as they tend to perform poorly if the individual features are not standard normally distributed, i.e. Gaussian with unit variance and zero means. Standardization is performed using Scikit-Learn's \cite{pedregosa11a} \texttt{StandardScalar} method. The output of the network is scaled back to the original distribution before computing the loss. 
    
The choice of the ANN architecture, such as the number of layers and the number of neurons in each layer, represents a trade-off between the quality of fit and the overfitting problem. The model should contain sufficiently many trainable parameters to learn the ground truth function accurately. At the same time, the ANN must avoid overfitting the training data by losing its ability to generalize well. As there is no panacea for choosing an optimal model configuration, the final architecture for the ANN was chosen based on empirical tests performed on the data and selecting the set of configurations that performed best on the validation set. Our network consists of two hidden layers with 15 neurons each and uses the rectified linear unit (ReLU) activation function, which is a standard choice for most deep learning applications and is known to mitigate the vanishing gradients problem \cite{Lecun2015}. The ReLU activation function behaves as an identity function for positive inputs and saturates at 0 for negative inputs. No activation function is applied to the input or the output layer. The design of our feedforward architecture is summarized in Table \ref{tab:ANN-arch}. 
    
The neural network is implemented and trained using the Python Keras library \cite{chollet2015keras} with a TensorFlow \cite{abadi2016tf} backend. Neural network parameters are initialized with the Glorot uniform distribution \cite{pmlr-v9-glorot10a}. We use the Adam optimizer \cite{kingma2014} to update the weights of the ANN.  For training, we use an initial learning rate of $ 1 \times 10^{-3}$ and the training data is batched into batches of size 16. Training is performed for a total of 40 epochs, or until the validation loss stops decreasing. The model is trained on a 2-core Intel Xeon CPU @ 2.20GHz. The network seeks to optimize a root mean squared error (RMSE) loss between the predictions and actual neutron star properties.

    \begin{table}[t]
        \centering
        \renewcommand{\arraystretch}{1.2}
        \begin{tabular}{ccc}
            \hline\hline
            \textbf{Layer} & \textbf{Number of neurons} & \textbf{Activation function}\\
            \hline
            0 (Input) & 7 & None\\
            1, 2 & 15 & ReLU\\
            3 (Output) & 6 & None\\
            \hline\hline
        \end{tabular}
        \caption{Neural network architecture used in this work. The input layer consists of seven neurons corresponding to nuclear matter parameters.
        In the output layer, the six neurons correspond to the six neutron star properties.}
        \label{tab:ANN-arch}
    \end{table}

\subsection{Bayesian statistical framework}
\label{sec:bayes}
    
The Bayesian statistical framework allows us to carry out a detailed analysis of the parameters of a model for a given set of fit data \cite{Wesolowski:2015fqa, Furnstahl:2015rha, Ashton2019}. The hypothesis or the prior knowledge of the model parameters and various constraints on them is encoded through the prior distributions. The technique yields a joint posterior distribution of the model parameters by updating the probability for the hypothesis using the available observational data according to Bayes' theorem. The posterior distribution of the model parameters $\boldsymbol\theta$ can be written as

\begin{equation}
    P(\boldsymbol \theta|D) = \frac{\mathcal{L}(D | \boldsymbol\theta) P(\boldsymbol\theta)}{\mathcal{Z}}
\end{equation}
where $\boldsymbol\theta$ and $D$ denote the set of model parameters and the fit data, respectively. $ P(\boldsymbol\theta| D)$ is the joint posterior probability of the parameters, $\mathcal{L}(D | \boldsymbol\theta)$ is the likelihood function, $P(\boldsymbol\theta)$ is the prior for the model parameters and $\mathcal{Z}$ is the evidence. The posterior distribution of a given parameter an be obtained by marginalizing $P(\boldsymbol \theta | D)$ over remaining parameters. The marginalized posterior distribution for a parameter $\theta_i$ is obtained as
    
\begin{equation}
    P(\theta_i | D) = \int P(\boldsymbol\theta | D) \prod_{j \neq i} d\theta_j
\end{equation}

A major contribution of this study is to demonstrate how a neural network can be effectively integrated within a Bayesian inference framework for astrophysical use cases. In particular, we demonstrate the effect of intercorrelations between symmetry energy parameters on NS properties. We sample NMPs within a Bayesian framework in two different ways: (i) sampling all the NMPs randomly without any intercorrelations among them, (ii) sample all the NMPs randomly, except for the pair $L_{\rm sym, 0}$-$K_{\rm sym, 0}$ which is set to a correlation coefficient of 0.9. Then, we compare the posterior distribution of the NMPs for both cases under certain constraints on NS properties.

For the sampling process, we use the Gaussian likelihood function defined as,
\begin{equation}
\mathcal{L}(D |  \boldsymbol\theta) = \prod_{j}\left(\frac{1}{\sqrt{2\pi}\sigma_{j}}\right)\exp\left[-\frac{1}{2}\sum_{i=1}^{N_d}\left( \frac{d_i - m_j(\boldsymbol\theta)}{\sigma_j} \right)^2\right]
\label{eq:likelihood}
\end{equation}
Here the index $j$ runs over all the data, $d_j$ and $m_j$ are the data and the corresponding model values, respectively. The model $m$ is parameterized by a set of parameters, $\boldsymbol\theta$. $\sigma_j$ are the adopted uncertainties. The evidence is used to compare the compatibility of different models with the available data. In our present work, the evidence $\mathcal{Z}$ is not relevant and thus can be ignored.
To obtain the marginalized posterior distributions of the NMPs within the Bayesian framework, we require a set of fit data, a model, and a set of priors for the nuclear matter parameters. The likelihood function for a given set of fit data is evaluated for a sample of NMPs populated according to their prior distributions. The joint probability distribution of the NMPs is obtained using the product of the likelihood function and the prior distribution. The fit data for the likelihood function is provided by the neural network for a sample drawn from the prior distribution. To compute the likelihood, we use the maximum neutron star mass $M_{\rm max}$, the maximum radius $R_{\rm max}$, the radius $R_{1.4}$, and the tidal deformability $\Lambda_{1.4}$, from the set of outputs generated by the neural network for a given input of NMPs. Instead of using a distinct value for each data point, we fix $d_i$ in Eq. (\ref{eq:likelihood}) to a mean value $\mu_j$. Therefore for our study, Eq. (\ref{eq:likelihood}) is modified to for a set of NMPs $\boldsymbol{p}$ and the neural network $\text{ANN}(\cdot)$ which accepts NMPs as inputs. We define the set of priors over the NMPs as a multivariable Gaussian distribution. The mean and standard deviation on each NMP in the distribution are listed in Table \ref{tab:priors}.
\subsubsection{The Log-Likelihood}
\label{sec:like}
The log-likelihood for mass, radius for 2.07 and 1.4 M$_\odot$ neutron star are defined as follows:
\begin{equation}
    log \chi(M_{max})=log \bigg[\frac{1}{exp\bigg[\frac{M_{cal}-M_{obs}}{-0.01}\bigg]+1} \bigg]
	\label{eq:masslog}
\end{equation}

\begin{equation}
    log \chi(R_{2.07})=-0.5 \bigg[ \frac {R_{2.07 cal}-R_{2.07 obs}}{\Delta R_{2.07}}\bigg]^2 + log\big[2 \pi \Delta R_{2.07}^2\big]
	\label{eq:r2.08log}
\end{equation}

\begin{equation}
    log \chi(R_{1.4})=-0.5 \bigg[ \frac {R_{1.4 cal}-R_{1.4 obs}}{\Delta R_{ 1.4}}\bigg]^2 + log\big[2 \pi \Delta R_{ 1.4}^2\big]
	\label{eq:r1.4log}
\end{equation}
The calculations are performed for a set of NS observational properties, such as maximum mass ($M_{max}$), the radius at NS mass ($R_{2.07}$), and at 1.4$M_{\odot}$ ($R_{1.4}$) listed in Table  \ref{tab:likelihood}. 

\begin{table}[]
\centering
\renewcommand{\arraystretch}{1.2}
\begin{tabular}{ccccc}
\hline \hline 
NMP                & \multicolumn{2}{c}{P1}                                           & \multicolumn{2}{c}{P2}                                           \\ \hline
$\boldsymbol{p_i}$ & $\boldsymbol{\mu_{p_i}}$ & $\boldsymbol{\sqrt{\Sigma_{p_ip_i}}}$ & $\boldsymbol{\mu_{p_i}}$ & $\boldsymbol{\sqrt{\Sigma_{p_ip_i}}}$ \\ \hline
$e_0$              & $-15.99$                 & $0.25$                                & $-15.99$                 & $0.50$                                \\
$\rho_0$           & $0.158$                  & $0.005$                               & $0.158$                  & $0.01$                                \\
$K_0$              & $242$                    & $15$                                  & $242$                    & $30$                                  \\
$Q_0$              & $-307$                   & $70$                                  & $-307$                   & $140$                                 \\
$J_{\rm sym,0}$    & $32.65$                  & $2.8$                                 & $32.65$                  & $5.6$                                 \\
$L_{\rm sym,0}$    & $73$                     & $12$                                  & $73$                     & $24$                                  \\
$K_{\rm sym,0}$    & $-20$                    & $50$                                  & $-20$                    & $100$                                 \\ \hline
\end{tabular}
\caption{The mean value $\boldsymbol{\mu_{p_i}}$ and error $\boldsymbol{\sqrt{\Sigma_{p_i p_i}}}$ for the nuclear matter parameters $p_i$ in the prior multivariate Gaussian distribution. All quantities are in units of MeV except for $\rho_0$ which is in units of fm$^{-3}$.}
\label{tab:priors}
\end{table}

Bayesian parameter estimation is commonly carried out using the Markov Chain Monte Carlo (MCMC) algorithm. This algorithm updates the existing parameters with a new set of parameters with a probability proportional to the ratio of the two points. However, MCMC approaches can lead to issues with converging to a stable posterior. To overcome this problem, we use the dynamic nested sampling algorithm \cite{Skilling2004,higson2019,speagle2020}. In dynamic nested sampling, the posterior is broken into many nested ``slices" with an initial \texttt{n-live} number of points that vary dynamically as the sampling progresses; samples are generated from each of them and then recombined to construct the posterior distribution.  We use the Dynesty dynamic nested sampling algorithm interfaced in BILBY \cite{Ashton2019} with 5000 \texttt{n-live} points to sample from the posterior distributions of the nuclear matter parameters. Dynesty enables flexible Bayesian inference over complex, multi-modal distributions without needing to converge to the posterior before generating valid samples \cite{speagle2020}.

\begin{table}
	\centering
	\caption{The constraints imposed in the Bayesian inference: Observed maximum mass of NS, Radius of 2.07 $M_{\odot}$ NS, Radius of 1.4 $M_{\odot}$ NS.}
	\label{tab:tab_const}
	\begin{tabular}{lccr} 
		\hline
		  &  \textbf{Constraints}&  &  \\
        Quantity & Value/Band & Reference\\
		\hline
		$M_{max}$ & $>$ 2.0 $M_{\odot}$ & \citep{Miller:2021qha}\\
		$R_{2.07}$ & $12.4 \pm 1.0$ km & \citep{Miller:2021qha}\\
		$R_{1.4}$ & $13.02 \pm 1.24$ km & \citep{Miller:2019cac}\\
		\hline
	\end{tabular}
 \label{tab:likelihood}
\end{table}

\section{Results}
\label{sec:results}

\begin{figure}[ht]
        \centering
        \includegraphics[scale=0.4]{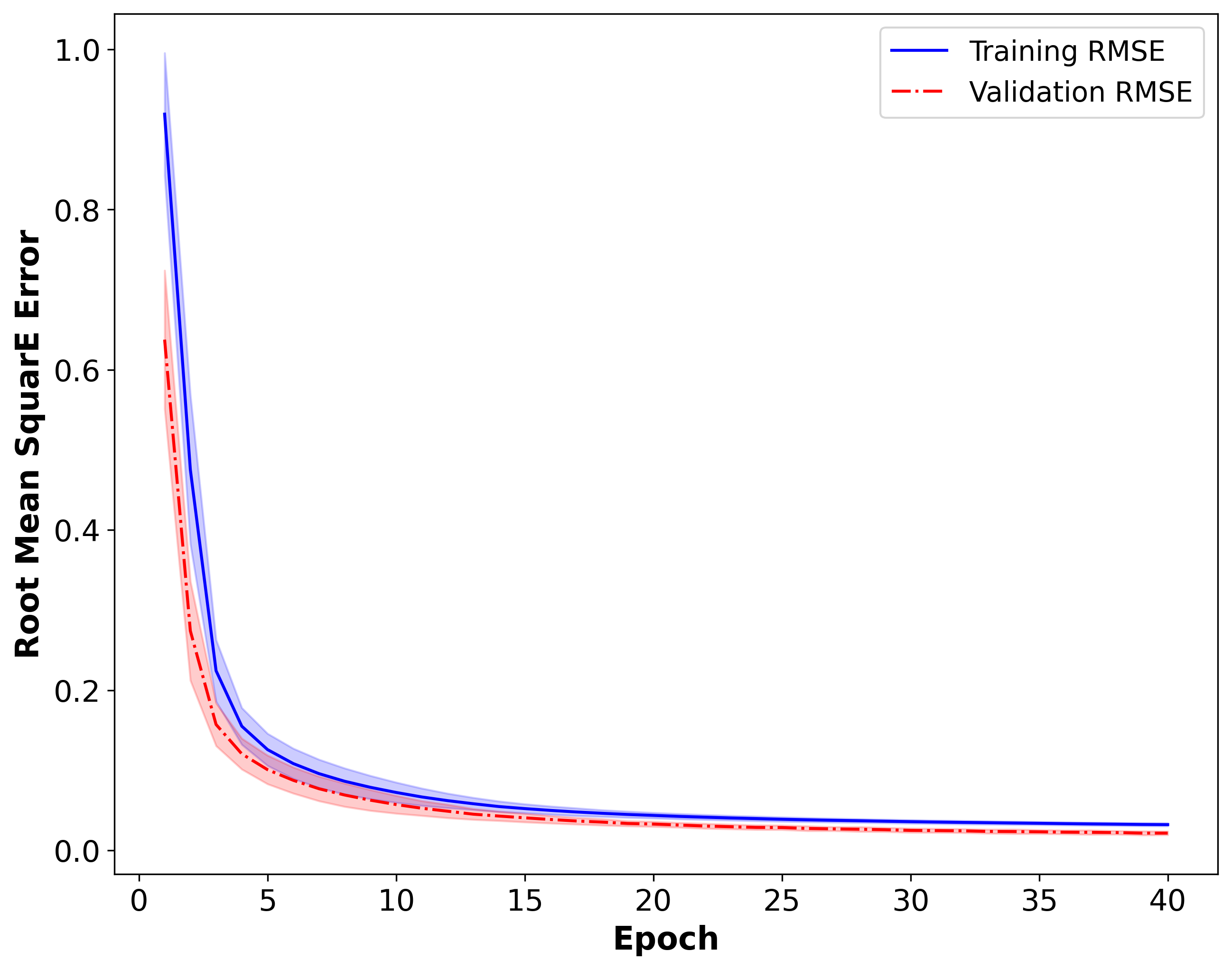}
        \caption{Learning curves for the training (blue, solid) and validation (red, dashed) data sets plotted against the number of elapsed epochs. The shaded region around the solid curve indicates a $1\sigma$ or 68\% CI region around the central value computed for 10 independent training runs.}
        \label{fig:losscurve}
    \end{figure}

In this section, we present the main findings of our analysis. After selecting the best model, which we call the NS-ANN, by performing a grid search over hyperparameters such as the model depth, model width, learning rate, etc., we determine its performance on the test set. We wish to reemphasize that the test set is never used during training or the validation phase. Evaluating the model's performance on this set quantifies the generalization capacity of the model, i.e., its predictive power on unseen data. The RMSE values obtained for each NS observable on the testing data set are summarized in Table \ref{tab:ann-metrics}. We also include the root mean squared relative error as it provides a scale-independent measure of the generalization capacity and eases comparison between multiple dependent variables. In Figure \ref{fig:losscurve}, we plot the loss as a function of training time, which is called the learning curve. The learning curve plots the root mean squared error on the $y$-axis against the number of elapsed epochs --- or, the number of full passes over the training data set --- on the $x$-axis. We plot two different learning curves for a single training instance: one for the training loss (in blue) and the other for the validation loss (in red). Figure \ref{fig:losscurve} also plots a $1\sigma$ (68\% confidence interval) region centered around the training curves computed for 10 independent runs to indicate the degree of variability observed by training the model on different subsets of the original data set. The training and validation losses are the loss functions computed for the training set and validation sets, respectively. The former is a metric of how well the ANN learns the training data, while the latter indicates how well the model is able to predict by generalizing the learned information. 
For a NS mass of 1.4$M_{\odot}$, we have a prediction error below 2\% for the radius $R_{1.4}$ and 5\% for the tidal deformability $\Lambda_{1.4}$. This implies that using the ANN, we can infer $\Lambda_{1.4}$ and $R_{1.4}$ with an average error of 19.239 and 0.194 km, respectively. The NS maximum mass, $M_{\rm max}$, and the maximum radius $R_{\rm max}$, are also predicted with an error below 2\%. Moreover, the saturation of the loss curves for the training and validation sets indicates that the training has converged to a minimum and that the model can generalize sufficiently well.

    \begin{table}[t]
        \centering
        \renewcommand{\arraystretch}{1.2}
        \begin{tabular}{ccc}
            \hline\hline
            $\boldsymbol{\hat{y}}$ & \textbf{RMSE} & \textbf{(RMSE/$\boldsymbol{\bar{y}}$) $\boldsymbol\times \boldsymbol{100}$}\\
            \hline
            $M_{\rm max}$ & 0.024 & 1.1\\
            $R_{\text{max}}$ & 0.088 & 0.8 \\
            $R_{1.4}$ & 0.194 & 1.5 \\
            $\Lambda_{1.0}$ & 123.800 & 3.7 \\
            $\Lambda_{1.4}$ & 19.239 & 4.3 \\
            $\Lambda_{1.8}$ & 5.344 & 8.2 \\
            \hline\hline
        \end{tabular}
        \caption{Root Mean Squared Error (RMSE) on the test set, defined as $\sqrt{(1/N)\sum_{i=1}^{N}(\hat{y}_i - y_i)^2 }$, where $N$ is the 
        a total number of samples in the test set. We also show the root mean squared relative error, (RMSE$/\bar{y}) \times 100$, where $\bar{y} = (1/N)\sum_{i=1}^{N}y_i$
        is the mean of the true value. All quantities are dimensionless except $R_{\rm max}$ and $R_{1.4}$, which are in units of km, and $M_{\rm max}$, which is in units of $M_{\odot}$.}
        \label{tab:ann-metrics}
    \end{table}
    
\begin{figure*}[!t]
    \centering
    \includegraphics[scale=0.4]{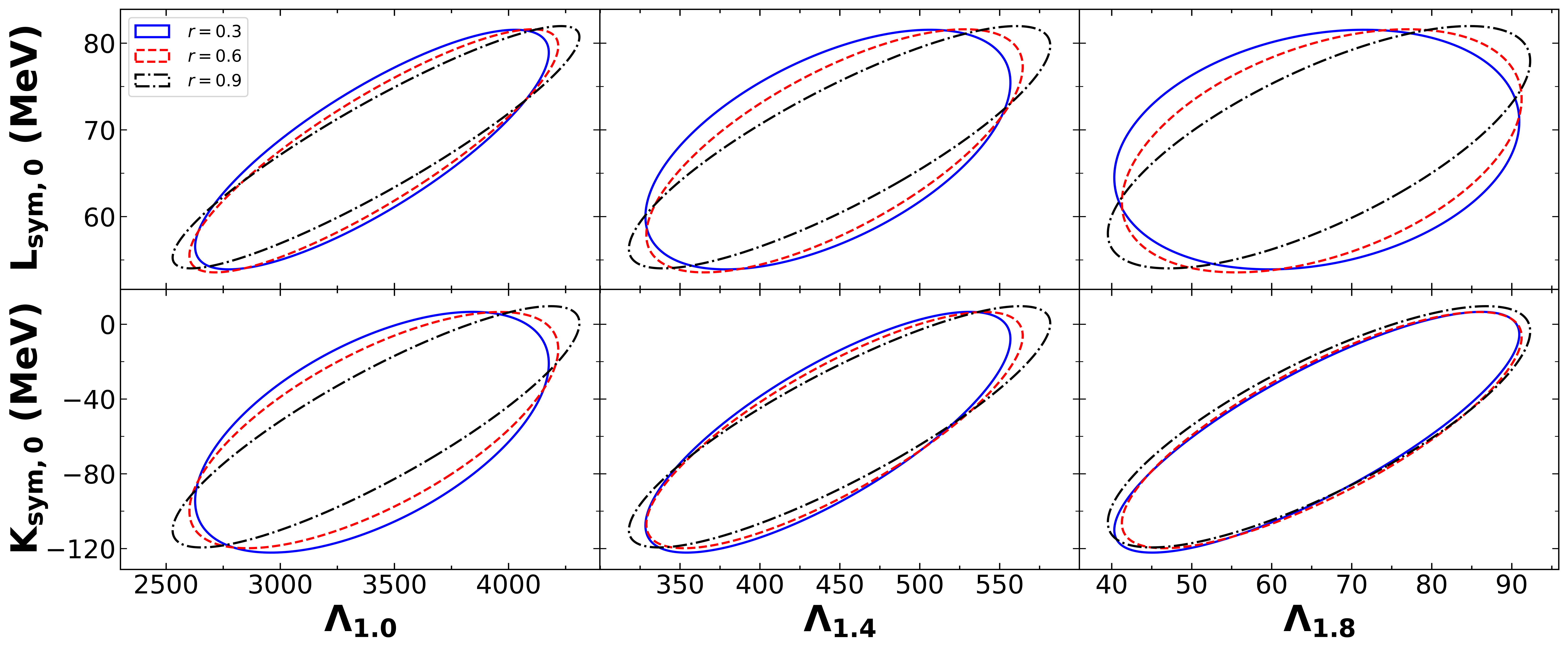}
    \caption{The 1$\sigma$ confidence ellipses in the planes of $\Lambda_M$ - $L_{\rm sym,0}$ (top) and $\Lambda_M$ - $K_{\rm sym, 0}$ (bottom) with $M = 1.0, 1.4$ and $1.8 M_{\odot}$ generated by the NS-ANN model for the correlation coefficient, $r$, between $L_{\rm sym,0}$ and $K_{\rm sym, 0}$ set to $0.3, 0.6$ and $0.9$.}
    \label{fig:corr-ellipse}
\end{figure*}

Once we have successfully and efficiently learned the non-linear mapping between the empirical nuclear matter parameters and the NS observables, we must first prove the correctness of the NS-ANN model before using it in the context of a Bayesian inference framework to constrain the NMPs. In Figure \ref{fig:corr-ellipse} we plot the $1\sigma$ confidence ellipses for  $\Lambda_M$ versus $L_{\rm sym,0}$ and $K_{\rm sym,0}$ for NS mass $M = 1.0,1.4,$ and $1.8 M_{\odot}$ predicted by the NS-ANN for three different pseudo-sets of NMPs. These pseudo-NMP distributions were constructed as a multivariate Gaussian distribution by setting the inter-correlation between $L_{\rm sym,0}$ and $K_{\rm sym, 0}$ to $r = 0.3, 0.6$ and $0.9$, and all other parameters of the distribution to values listed in Table \ref{tab:nmp-params}. Once again, all NMPs that generate EoSs which do not satisfy the conditions from Sec. \ref{sec:anns} are discarded. We use the NS-ANN model for each NMP in the dataset to generate a corresponding set of NS observables. Through this analysis, we wish to establish that a neural network is capable of replicating the underlying microphysical information conveyed by a conventional EoS model. The values of the correlation coefficients for the results illustrated in the figure are summarized in Table \ref{tab:corr-results}. For the first set, we observe that the correlations of $\Lambda_{1.0, 1.4, 1.8}$ with $K_{\rm sym, 0}$ are $\chi \sim 0.5 - 0.8$, and those with $L_{\rm sym, 0}$ are $\chi \sim 0.2 - 0.8$. For the second set, we see a non-trivial narrowing of the confidence ellipses, indicating stronger correlations of $\Lambda_{1.0}$ and $\Lambda_{1.4}$ with  $L_{\rm sym, 0}$ and $K_{\rm sym, 0}$, $\chi \sim 0.7 - 0.8$, while these correlations become moderate for $\Lambda_{1.8}$. We also observe that the $\Lambda_M - L_{\rm sym, 0}$ correlations decrease with increasing mass of NS, while an opposite trend is observed for $\Lambda_M - K_{\rm sym, 0}$ correlations. Moreover, for the same NS mass, $\Lambda_M - L_{\rm sym, 0}$ correlations are much more sensitive to $ L_{\rm sym, 0} - K_{\rm sym, 0}$ correlations over $\Lambda_M - K_{\rm sym, 0}$ correlations. These results help emphasize that the correlations of the tidal deformability with $L_{\rm sym, 0}$ and $K_{\rm sym, 0}$ are sensitive to the physical correlations among the empirical NMPs arising from a set of physical constraints \cite{Margueron:2017eqc, Margueron:2017lup}. Additionally, the observations made in this analysis are in excellent agreement with results from previous studies performed on observing the effect of correlations of the tidal deformability with the slope of the symmetry energy, and its curvature \cite{Malik:2020vwo, Fattoyev:2017jql, Ferreira:2019bgy}. At this point we also wish to note that there exists a strong model-independent intercorrelation between $L_{\rm sym,0}$ and $J_{\rm sym,0}$, but this correlation does not disrupt the relationships between some NS properties, such as tidal deformability, radius, and NMPs \cite{Malik:2020vwo}, and as a result is not covered in this study.
 
\begin{table}[htbp]
    \centering
    \renewcommand{\arraystretch}{1.2}
    \begin{tabular}{ccccc}
        \hline\hline
        $r$ & & $\Lambda_{1.0}$ & $\Lambda_{1.4}$ & $\Lambda_{1.8}$\\
        \hline
        \multirow{2}{*}{0.3} & $L_{\rm sym, 0}$ & 0.81 & 0.56 & 0.23\\
                            & $K_{\rm sym, 0}$ & 0.57 & 0.78 & 0.81\\
        \hline
        \multirow{2}{*}{0.6} & $L_{\rm sym, 0}$ & 0.86 & 0.69 & 0.43\\
                            & $K_{\rm sym, 0}$ & 0.69 & 0.80 & 0.79\\
        \hline
        \multirow{2}{*}{0.9} & $L_{\rm sym, 0}$ & 0.91 & 0.84 & 0.71\\
                            & $K_{\rm sym, 0}$ & 0.86 & 0.86 & 0.79\\
        \hline\hline
    \end{tabular}
    \caption{The values of the correlation coefficients for $\Lambda_{1.0, 1.4, 1.8}$ with $L_{\rm sym, 0}$ and $K_{\rm sym, 0}$ for three different $L_{\rm sym, 0}-K_{\rm sym, 0}$ correlation coefficients, $r$}
    \label{tab:corr-results}
\end{table}

As a next step, we apply our trained NS-ANN model to a dataset acquired from a different class of physics models in order to assess the model-dependent uncertainty related to neutron star properties. 

The nuclear matter parameters are known to be model-dependent and hence also the properties of stars or EoSs \cite{Pradhan:2022txg,Malik:2023mnx,Zhou:2023hzu}. 
In Table \ref{tab:ddb}, we compare the NS properties obtained using a dataset generated by a density-dependent meson coupling framework called the DDB model. This model is constructed using a Bayesian inference approach in conjunction with minimal constraints on the nuclear saturation properties, the maximum mass of a neutron star exceeding $2M_\odot$, and a low-density equation of state (EoS) calculated using chiral effective field theory. In the first horizontal panel, we show that NS properties predicted by the NS-ANN ML model match well with the theoretical predictions of the Skyrme model. In the second horizontal panel, then we apply our trained NS-ANN model to a data set generated by DDB. It can be seen that the residual errors on the NS properties are much larger in this case. We posit that this wider uncertainty might arise due to input parameters (NMPs) in the DDB dataset lying outside the parameter space covered by the training set. To avoid any extrapolation error, we then filter the DDB according to the parameter space of the training sets. We call this set DDB-FL and apply the NS-ANN on this set. It can be seen that the dispersion is still large. The NS-ANN model is trained on data obtained from the non-relativistic Skryme model, so it is unsurprising that it fails to completely capture the features of the relativistic DDB model. 

\begin{table*}[]
\setlength{\tabcolsep}{6.5pt}
      \renewcommand{\arraystretch}{1.2}
\begin{tabular}{ccccc|cccccccc}
\hline \hline 
\multicolumn{5}{c|}{Input}                                                                         & \multicolumn{7}{c}{Output}                                                                                                           & \multicolumn{1}{l}{}                     \\ \cline{1-13}
\multicolumn{2}{c}{\multirow{2}{*}{NMP}}  & \multirow{2}{*}{median} & \multicolumn{2}{c|}{90\% CI} & \multirow{3}{*}{NSP} & \multicolumn{3}{c}{Skyrme}                            & \multicolumn{3}{c}{ML}                                & \multicolumn{1}{l}{$\sigma_{rr}$} \\ \cline{4-5} \cline{7-12}
\multicolumn{2}{c}{}    &                         & min           & max          &                      & \multirow{2}{*}{median} & \multicolumn{2}{c}{90\% CI} & \multirow{2}{*}{median} & \multicolumn{2}{c}{90\% CI} & \multicolumn{1}{l}{}                     \\ \cline{1-5}
\multirow{7}{*}{Skyrme} & $e_0$           & $-16.01$                & $-16.39$      & $-15.59$     &                      &                         & min          & max          &                         & min          & max          & \multicolumn{1}{l}{}                     \\ \cline{6-13} 
                        & $\rho_0$        & $0.160$                 & $0.152$       & $0.168$      & $M_{\rm max}$        & $2.09$                  & $1.87$       & $2.24$       & $2.08$                  & $1.89$       & $2.24$       & $2.2$                                    \\
                        & $K_0$           & $231$                   & $200$         & $261$        & $R_{\rm max}$        & $10.74$                 & $9.77$       & $11.85$      & $10.75$                 & $9.80$       & $11.84$      & $1.5$                                    \\
                        & $Q_0$           & $-292$                  & $-457$        & $-140$       & $R_{\rm 1.4}$        & $12.82$                 & $11.56$      & $14.10$      & $12.81$                 & $11.60$      & $14.02$      & $2.1$                                    \\
                        & $J_{\rm sym,0}$ & $32$                    & $27$          & $37$         & $\Lambda_{1.0}$      & $3279$                  & $1994$       & $5323$       & $3279$                  & $1973$       & $5280$       & $6.8$                                    \\
                        & $L_{\rm sym,0}$ & $67$                    & $42$          & $94$         & $\Lambda_{1.4}$      & $435$                   & $234$        & $729$        & $430$                   & $232$        & $722$        & $7.0$                                    \\
                        & $K_{\rm sym,0}$ & $-63$                   & $-169$        & $73$         & $\Lambda_{1.8}$      & $63$                    & $21$         & $118$        & $63$                    & $22$         & $116$        & $15.5$                                   \\
                        &                 &                         &               &              &                      &                         &              &              &                         &              &              & \multicolumn{1}{l}{}                     \\ \hline
\multirow{7}{*}{DDB}    & $e_0$           & $-16.10$                & $-16.41$      & $-15.80$     &                      & \multicolumn{3}{c}{DDB}                               & \multicolumn{3}{c}{ML}                                & \multicolumn{1}{l}{$\sigma_{rr}$} \\ \cline{7-13} 
                        & $\rho_0$        & $0.153$                 & $0.147$       & $0.158$      & $M_{\rm max}$        & $2.14$                  & $2.02$       & $2.35$       & $2.21$                  & $1.95$       & $2.39$       & $9.4$                                    \\
                        & $K_0$           & $232$                   & $201$         & $275$        & $R_{\rm max}$        & $11.33$                 & $10.97$      & $11.81$      & $10.54$                 & $9.83$       & $11.27$      & $11.3$                                   \\
                        & $Q_0$           & $-108$                  & $-256$        & $129$        & $R_{\rm 1.4}$        & $12.63$                 & $12.07$      & $13.22$      & $12.26$                 & $11.35$      & $13.46$      & $7.7$                                    \\
                        & $J_{\rm sym,0}$ & $32$                    & $29$          & $35$         & $\Lambda_{1.0}$      & $2802$                  & $2171$       & $3702$       & $2829$                  & $1869$       & $3911$       & $16.5$                                   \\
                        & $L_{\rm sym,0}$ & $41$                    & $24$          & $65$         & $\Lambda_{1.4}$      & $455$                   & $339$        & $626$        & $385$                   & $247$        & $582$        & $34.7$                                   \\
                        & $K_{\rm sym,0}$ & $-116$                  & $-150$        & $-73$        & $\Lambda_{1.8}$      & $79$                    & $52$         & $125$        & $70$                    & $28$         & $110$        & $65.4$                                   \\
                        &                 &                         &               &              &                      &                         &              &              &                         &              &              & \multicolumn{1}{l}{}                     \\ \hline
\multirow{7}{*}{DDB-FL} & $e_0$           & $-16.12$                & $-16.43$      & $-15.83$     &                      & \multicolumn{3}{c}{DDB-FL}                            & \multicolumn{3}{c}{ML}                                & \multicolumn{1}{l}{$\sigma_{rr}$} \\ \cline{7-13} 
                        & $\rho_0$        & $0.155$                 & $0.152$       & $0.159$      & $M_{\rm max}$        & $2.07$                  & $2.01$       & $2.16$       & $2.05$                  & $1.93$       & $2.18$       & $6.6$                                    \\
                        & $K_0$           & $216$                   & $202$         & $225$        & $R_{\rm max}$        & $11.13$                 & $10.94$      & $11.30$      & $10.32$                 & $9.90$       & $10.97$      & $10.5$                                   \\
                        & $Q_0$           & $-188$                  & $-260$        & $-144$       & $R_{\rm 1.4}$        & $12.43$                 & $12.12$      & $12.82$      & $12.37$                 & $11.81$      & $13.40$      & $4.7$                                    \\
                        & $J_{\rm sym,0}$ & $32$                    & $29$          & $35$         & $\Lambda_{1.0}$      & $2612$                  & $2271$       & $3379$       & $2678$                  & $2146$       & $3880$       & $12.6$                                   \\
                        & $L_{\rm sym,0}$ & $51$                    & $43$          & $72$         & $\Lambda_{1.4}$      & $396$                   & $337$        & $486$        & $343$                   & $257$        & $497$        & $28.7$                                   \\
                        & $K_{\rm sym,0}$ & $-123$                  & $-152$        & $-70$        & $\Lambda_{1.8}$      & $62$                    & $49$         & $76$         & $47$                    & $27$         & $75$         & $64.1$                                   \\ \hline
\end{tabular}
\caption{The median and the minimum (min) and maximum (max) values of the associated 90\% confidence intervals for the NMPs and the NS properties obtained from Skyrme, DDB, and DDB-FL datasets. We also indicate these quantities for maximum relative residual errors $\sigma_{rr}$ on the NS properties.}
\label{tab:ddb}
\end{table*}

\subsection{Validity of NS-ANN model in Bayesian Inference}
In this section, we apply our NS-ANN model in a Bayesian inference framework. The primary objective is to validate the trained model's effectiveness in computationally expensive calculations of this type. The results will be analyzed in two parts: i) The posterior of neutron star properties will be compared between trained NS-ANNs and Skyrme models in Bayesian Inference, ii) the CPU time will be compared between the two models.

\subsection{Comparison between NS-ANN and Skyrme}
As discussed above, Bayesian inference requires three ingredients: a) the model, b) the fit data or likelihood, and c) prior. We will perform identical Bayesian inference calculations with both Skyrme and NS-ANN models and compare their posteriors. The input priors are the nuclear matter parameters, namely $e_0$, $\rho_0$, $K_0$, $Q_0$, $J_{\rm sym,0}$, $L_{\rm sym,0}$, and $K_{\rm sym,0}$. In Table \ref{tab:priors}, we have defined two multivariate Gaussian priors, P1 and P2. P1 represents the prior validity in which the neural network is originally trained, while in P2, we have doubled the uncertainty to evaluate the performance of the NS-ANN model. The likelihood for our calculation is defined in section \ref{sec:like} for the fit data described in \ref{tab:likelihood}. In the likelihood, we have only considered observed NS properties because we are motivated to compare the performance of the NS-ANN trained model with the Skyrme nuclear physics model within observational constraints.

\begin{figure*}
        \centering
        \includegraphics[scale=0.45]{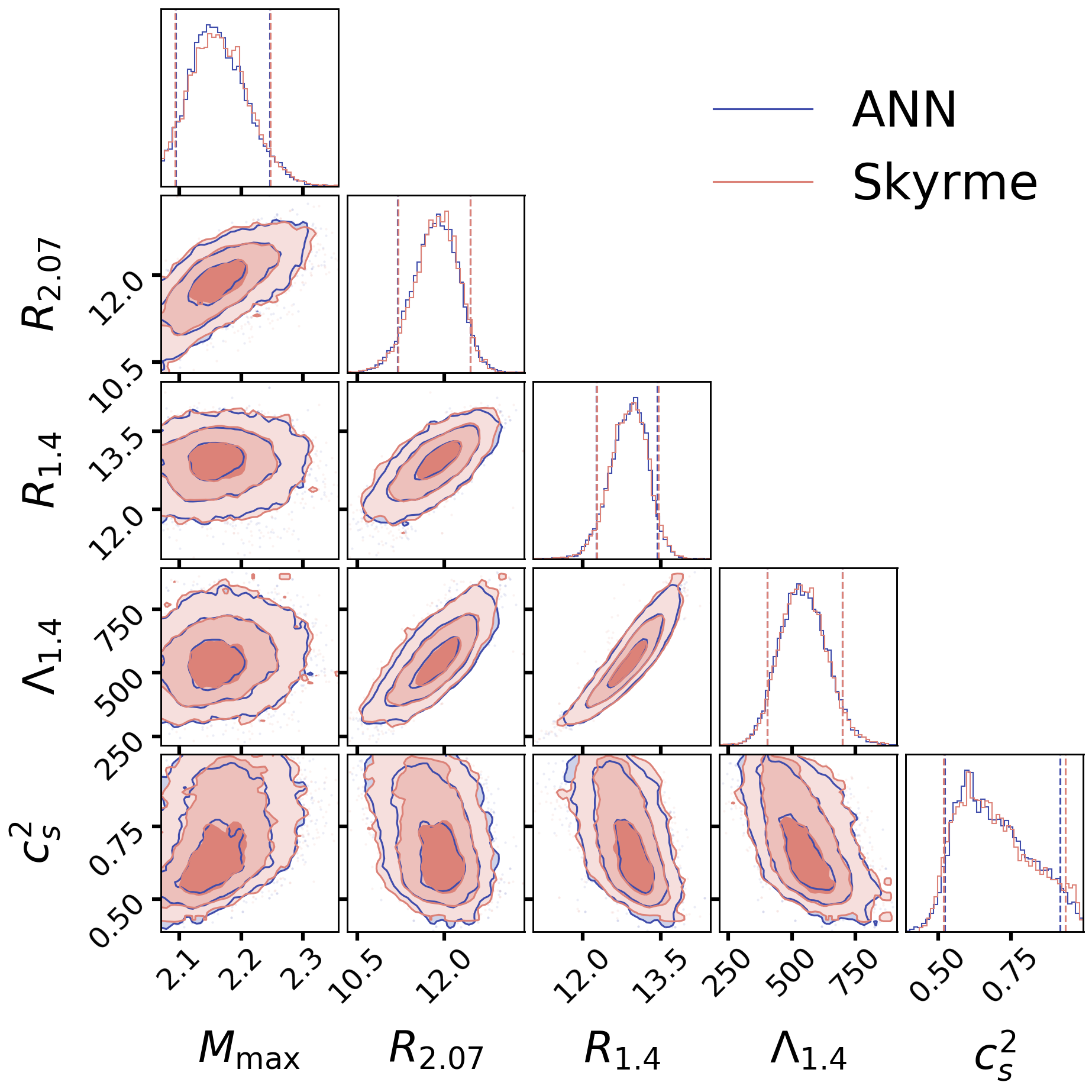}
        \caption{Comparison of neutron star properties obtained from Bayesian posterior for prior set P1 with ANN and Skyrme model, including NS maximum mass ($M_{\rm max}$), radius for a 2.07 M$_\odot$ NS, radius ($R{_1.4}$) and tidal deformability ($\Lambda_{1.4}$) for a 1.4 M$_\odot$ NS, and square of the speed of sound ($C_s^2$) at the center of the maximum mass NS. The vertical lines represent the 68\% CIs, and the light and dark intensities represent the 1$\sigma$, 2$\sigma$, and 3$\sigma$ CIs, respectively.}
        \label{fig:corner-plot_c1}
\end{figure*}

\begin{figure*}
        \centering
        \includegraphics[scale=0.45]{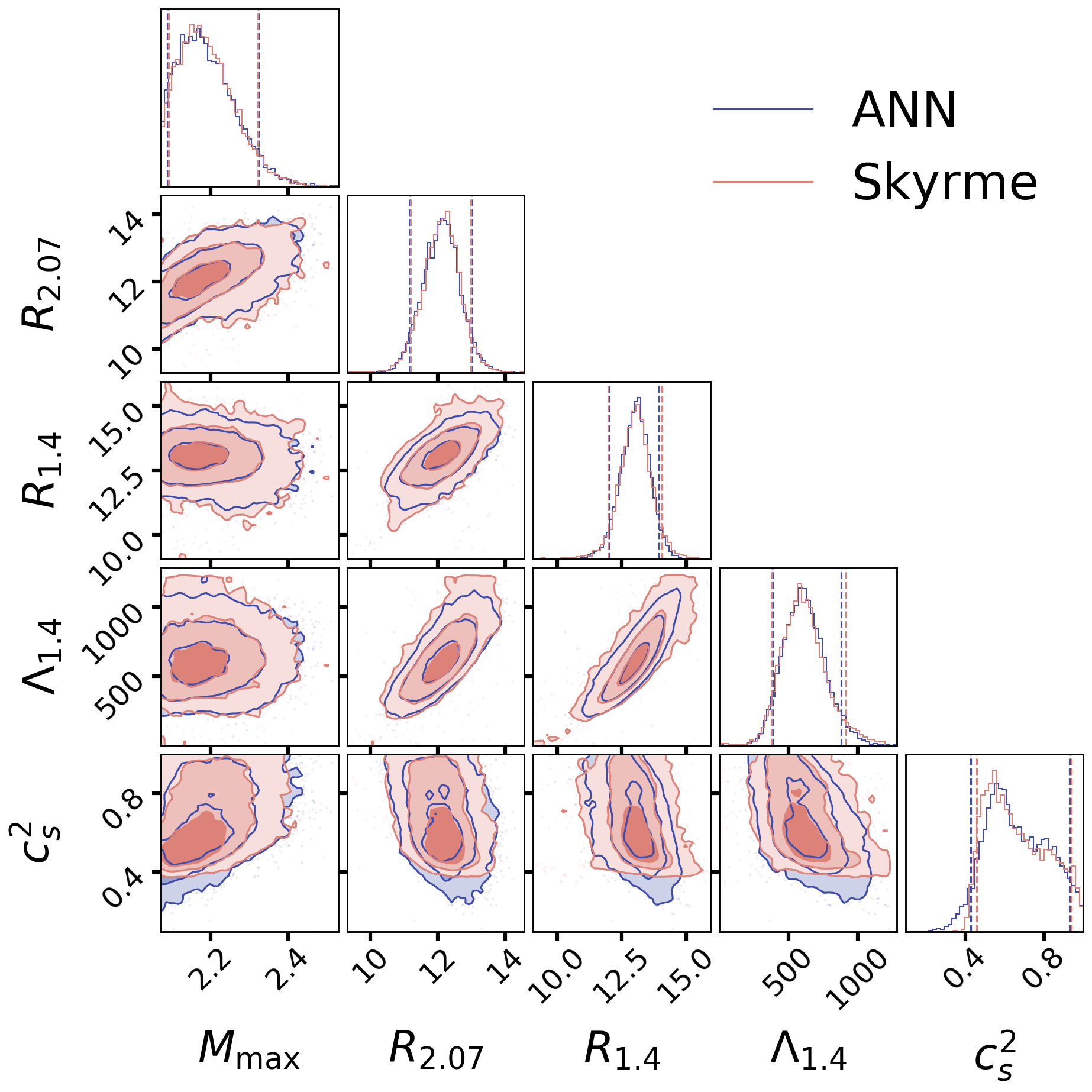}
        \caption{Same as figure \ref{fig:corner-plot_c1} but for prior set P2.
        }
        \label{fig:corner-plot_c2}
\end{figure*}

We used the Pymultinest sampler to sample our parameter. In Figure \ref{fig:corner-plot_c1}, we compared the Bayesian posterior results for the neutron star properties obtained from the prior set P1 with the NS-ANN and Skyrme model. This included the NS maximum mass (Mmax), radius for a 2.07 M$\odot$ NS, radius (R$_{1.4}$), tidal deformability ($\Lambda_{1.4}$) for a 1.4 M$_\odot$ NS, and the square of the speed of sound ($C_s^2$) at the center of the maximum mass NS. We also compared the same for the prior set P2 in Figure \ref{fig:corner-plot_c2}. In both figures, the joint distribution of NS properties is represented by corner plots, which are obtained from Bayesian Inference posteriors. Each variable in a corner plot is plotted on one of the axes, and its distribution is shown by its diagonal. Off-diagonal plots show the joint distributions of each pair of variables. Using it to understand how variables are related is a useful tool.  Now, it's time to compare their accuracy and validity.

In Table \ref{tab:model-rmse}, we compared the relative \% residuals of the NS-ANN model with the Skyrme model for both P1 and P2. It is worth noting that the results for P1 are very comparable to the original physics model. This is expected because P1 is the prior range in which the NS-ANN was originally trained. The relative \% residuals for the NS maximum mass, radius for 2.07 and 1.4 M$\odot$ NS, and tidal deformability for 1.4 M$\odot$ NS are $\sim1\%$, while the square of the speed of sound is $\sim3\%$ at the center of the maximum mass NS. However, in the case of P2, all the residuals are noticeably increased, with the tidal deformability of 1.4 M$_\odot$ NS and the square of the speed of sound at the center of the maximum mass NS being around $\sim5\%$. It should be emphasized that the neural network model can be trusted for interpolation but used with caution for extrapolation. The main advantage of this type of ANN model is that it can produce very fast draft results for computationally expensive calculations.

\subsection{CPU Time}
In Table \ref{tab:model-times}, we compared the CPU time for NS-ANN and Skyrme model for both prior sets. The timing tests were performed on a 12-core Intel i7-8700K CPU @ 3.70 GHz. It is noteworthy that the NS-ANN model is approximately 500 times faster than the Skyrme model. It is exciting that the neural network model can be employed in computationally expensive calculations to obtain fast draft results. This will help in reducing carbon emissions and help reduce the environmental cost of using high computational power for the preliminary testing of any model. 
\begin{table}[]
\centering
\begin{tabular}{ccccccc}
\hline \hline 
\multicolumn{2}{c}{NS}                             & M$_{\rm max}$ & R$_{\rm 2.07}$ & R$_{\rm 1.4}$ & $\Lambda_{\rm 1.4}$ & C$_s^2$ \\
\multicolumn{2}{c}{Units}                          & M$_\odot$     & km             & km            & ...                 & c$^2$   \\ \hline
\multirow{2}{*}{Maximum relative \% residual} & P1 & 0.3           & 0.6            & 0.3           & 1.2                 & 2.5     \\
                                              & P2 & 0.8           & 1.6            & 1.5           & 5.3                 & 5.1     \\ \hline
\end{tabular}
\caption{The maximum relative residual for the maximum mass of a neutron star (NS), denoted as M${\rm max}$, is attained. Additionally, we determine the radius R${2.07}$ for a 2.07 solar mass (M$\odot$) NS, the radius R${1.4}$ and the dimensionless tidal deformability $\Lambda_{1.4}$ for a 1.4 M$_\odot$ NS, as well as the squared speed of sound at the core of the NS with maximum mass. These quantities are obtained using the prior sets P1 and P2.}
\label{tab:model-rmse}
\end{table}

    \begin{table}[t]
        \centering
        \renewcommand{\arraystretch}{1.2}
        \begin{tabular}{ccc}
            \hline\hline
            \textbf{Model} & \textbf{P1} & \textbf{P2}\\
            \hline
            ANN & 2.23 min & 3.08 min \\
            Skyrme & 16h 27 min & 19h 55 min \\
            \hline\hline
        \end{tabular}
        \caption{CPU inference time estimates for the ANN model and a Skyrme model to infer NS observations from a set of NMPs for prior sets P1 and P2. The timing tests were performed on a 12-core Intex i7-8700K CPU @ 3.70 GHz. The inference is performed with a batch size of one.  }
        \label{tab:model-times}
    \end{table}

\section{Conclusions}
\label{sec:conclusion}
We have demonstrated the application of ANNs to analyze NS observables like the radius, mass, and tidal deformability from a set of seven parameters, or nuclear saturation properties that characterize the equation of state of cold dense matter. We have shown that a neural network that models such a mapping is able to learn the microphysics information of finite nuclei, which are the intercorrelations arising between the NMPs. Moreover, we delved into the utilization of Artificial Neural Networks (ANN) as a complement to theoretical models, enabling them to undertake computationally demanding tasks, such as Bayesian Inference studies to constrain nuclear models. {This study demonstrates that neural networks or some advanced AI tools not only function as efficient surrogates for traditional physics-based EoS models but also have the potential to eventually replace phenomenological models, particularly for computationally intensive or high-dimensional problems.} 

Let us briefly summarize what we have done in this work. First, we generate a set of pseudo-NMP data using the Skyrme model. A suitable set of NMPs are chosen so that the resulting neutron star EoSs are consistent with the currently observed maximum mass of $\sim 2M_{\odot}$ and satisfy causality constraints. Next, we train an artificial neural network on this data set to learn a non-linear mapping between the set of nuclear matter parameters and NS observables. We demonstrate that the ANN is capable of inferring, with reasonable accuracy, NS observables from empirical parameters, as compared to conventional physics models which require the computation of an EoS followed by tedious equation-solving. This NS-ANN model is then validated to ensure that it satisfies the microphysics information of finite nuclear matter. More specifically, we study whether a trained ANN model is able to capture correlations between the tidal deformability for a NS with mass $1.4M_{\odot}$, $\Lambda_{1.4}$ and the symmetry energy slope $L_{\rm sym,0}$ as well as its curvature $K_{\rm sym,0}$. We find that the NS-ANN model learns a mapping that is sensitive to $L_{\rm sym,0} - K_{\rm sym,0}$ correlations in agreement with previous studies. Using the Bayesian inference framework, we determine the extent to which the neural network can replicate the physics model, and then utilize the trained neural network for inference tasks, which typically involve computationally expensive calculations. By doing so, we are able to achieve results that are comparable to the physics model while significantly reducing the computational time required. 

Presently, our framework is a proof-of-concept that demonstrates the applicability of ANNs to NS physics. For a more realistic application of our framework, empirical uncertainties ought to be considered. This can be achieved, for example, by using a class of ANNs called Bayesian Neural Networks to perform the inference task. Bayesian networks have the ability to cast the problem into a probabilistic domain by inferring probability distributions over a prior of NMP inputs. A model which predicts uncertainties will also be able to potentially further reveal the effects of NS observational constraints on the NMPs. Presently, the entire domain of validity of the ANN cannot be automatically ensured. Due to a lack of training data for some parts in the modeled output space, the ANN might be unable to accurately predict regions that have not been learned. 

{While the current study employs traditional methods and neural network models for Bayesian analysis, we recognize the emerging potential of more sophisticated techniques. Specifically, Normalizing Flows and Convolutional Variational Auto-Encoders (CVAEs) offer promising avenues for future work. Normalizing flows and convolutional variational auto-encoders (CVAEs) are powerful methods able to model complex probability distributions \cite{Morrow2020}. Although normalizing flows provides an accurate evaluation of likelihood, they can be computationally demanding, especially when dealing with high-dimensional data. On the other hand, CVAEs are efficient and can handle large datasets, but only offer approximate likelihood evaluations. They may be useful when handling grid-like data, such as images. The decision to use either method depends on task requirements, such as exact likelihood estimation and computational efficiency. }

This work is limited only to hadronic NS compositions, i.e, the set of $\beta$-equilibriated EoSs employed in this work is composed of neutrons, protons, electrons, and muons. Present observational constraints on NS properties cannot rule out the possibility of exotic degrees of freedom or deconfined quark phases inside the NS core. 

{In future work, a detailed and systematic analysis with an ANN trained on a set of EoSs with different compositions of particles is required to investigate the uncertainties on higher-order NMPs with available observational constraints. The exploration of more advanced AI methods, such as Normalizing Flows and CVAEs, not only holds the promise for enhancing computational efficiency but also raises the possibility of becoming the de facto method for certain types of nuclear and astrophysical analyses.}

\section*{Data availability} 
The trained network and corresponding data sets are made publicly available through a GitHub repository. \footnote{\url{https://github.com/ameya1101/NS-ANN}}

\section*{Acknowledgements}
T.M. would like to thank national funds from FCT (Fundação para a Ciência e a Tecnologia, I.P, Portugal) under Projects No. UID/\-FIS/\-04564/\-2019, No. UIDP/\-04564/\-2020, No. UIDB/\-04564/\-2020, and No. POCI-01-0145-FEDER-029912 with financial support from Science, Technology, and Innovation, in its FEDER component, and by the FCT/MCTES budget through national funds (OE). The authors acknowledge the Laboratory for Advanced Computing at the University of Coimbra for providing {HPC} resources that have contributed to the research results reported within this paper.


\end{document}